# Experimental investigations of quasi-coherent micro-instabilities in J-TEXT Ohmic plasmas

Peng Shi[1,3], J.C. Li[2], * G. Zhuang[4], Zhifeng Cheng[3], Li Gao[3], Yinan Zhou[3]

[1] *United Kingdom Atomic Energy Authority, Culham Science Centre, Abingdon, Oxon OX143DB, United Kingdom*

[2] *Department of Earth and Space Sciences, Southern University of Science and Technology, 518055 Shenzhen, Guangdong, People's Republic of China*

[3] *State Key Laboratory of Advanced Electromagnetic Engineering and Technology, School of Electrical and Electronic Engineering, Huazhong University of Science and Technology, Wuhan 430074, People's Republic of China*

[4] *School of Physical Sciences, University of Science and Technology of China, Hefei, Anhui 230026, People's Republic of China*

*Corresponding author: lijc@sustech.edu.cn

The ITG and TEM instabilities with quasi-coherent spectra have been identified experimentally, by the newly developed far-forward collective scattering measurements in J-TEXT tokamak Ohmical plasmas. The ITG mode has characteristic frequencies in the range of $30 - 100 kHz$ and wavenumber of $k_\theta \rho_s < 0.3$. After the plasma density exceeds at critical value, the ITG mode shows a bifurcation behavior, featured by frequency decrease and amplitude enhancement. Meanwhile, the ion energy loss enhancement and confinement degradation are also observed. It gives the direct experimental evidence for ion thermal transport caused by ITG instability.

It is widely accepted that the anomalous transport rise from micro-instabilities or turbulence is the main mechanism for cross-field particle and heat transport in tokamaks [1,2]. Therefore, understanding the micro-instabilities in tokamaks is crucial for future fusion devices. For Ohmically heated tokamak plasmas, one of the most important micro-instabilities modes is the ion-temperature-gradient driven drift wave instabilities (ITG mode) [3,4]. Theory has long predicted that the ITG mode is the dominant microscopic turbulence and the dominant source of anomalous ion transport in tokamak plasmas [5,6]. However, there only exists sparse direct or indirect experimental evidences to implicate the ITG mode in particular instability in a tokamak [7,8]. So the dominant role of the ITG mode predicted by theory has not been firmly established experimentally in tokamaks. The challenge to directly distinguish ITG mode in tokamak plasmas is identifying the propagation direction for a specific turbulence. Because the ITG mode usually coexists with the trapped electron mode (TEM) and they have the similar wavelength scale such that $k_\theta \rho_s < 1$, where $k_\theta$ is the poloidal wave number and $\rho_s = \sqrt{m_i T_e}/(Z_i B)$ is the main ion Larmor radius with respect to the main ion sound speed. In early years, by use of far-infrared (FIR) collective scattering measurements, a turbulence with ion feature (propagated in the ion diamagnetic drift direction) was observed in the saturated Ohmic confinement (SOC) plasmas, which was referred to the ITG mode turbulence [9,10]. In recent decades, on the benefits of development of reflectometer diagnostics, a particular kind of density fluctuations called quasi-coherent (QC) modes con-cerning the ITG and TEM modes were widely observed in tokamaks [11-13]. The first studies concerning QC modes were performed in T-10 tokamak, which was reported in Ref. 11. They found two different QC fluctuations: low frequency (LF) QC and high frequency (HF) QC modes. By comparing with simulations, the LF QC mode was inferred as ITG instability while the HF QC mode was linked with TEM instability. Subsequent studies related to QC modes in TEXTOR and Tore Supra mainly focused on the LF QC modes due to the HF QC modes were not detected. More recently, a QC mode similar to the LF QC mode in T-10 was also observed on HL-2A and J-TEXT tokamaks by reflectometer [14]. But in contrast to the Ref. 11, authors of Ref. 12 &14 inferred that LF QC modes as the TEM instabilities, although they have quite similar characteristic frequency of $50 - 120 kHz$ and wave-number of $k_\theta \rho_s \cong 0.1 - 0.4$. The difference mainly arises from the lack of direct measurement for the propagation direction, which is the key point for judging the QC modes be ion or electron mode. In this sense, the FIR collective scattering measurements [9] have an advantage over reflectometer. But on the other hand, the collective scattering cannot identify the QC turbulences because it only measure the fluctuation with particular wave-number $k_\perp$. In a word, there still have not direct evidences to affirm the ITG or TEM modes in tokamak experiments.

Most recently, by using the newly developed far-forward collective scattering (FCS) measurements [15], the QC density fluctuation reported in Ref. 14 has also been detected and studied on Joint-TEXT tokamak (formerly TEXT-U), which is a conventional medium-sized tokamak with a major radius of $R_0 = 1.05\ m$ and minor radius of $a = 0.25m{\sim}0.29m$ (set by the silicon-carbide coated graphite limiter) [16]. The FCS measurement is based on the 17-channel three-wave FIR polarimeter-interferometer system (POLARIS) [17], which has the vertical impact parameters as $r = -24 : 3 : 24 cm$, where $r = R - R_0$. Here, $r < 0$ and $r > 0$ corresponds to high field side (HFS) and low field side (LFS) respectively. Additionally, the FCS measures the line-integral electron density fluctuations with wave-number in the range of $k_\perp < 1.5 cm^{-1}$, where the index ⊥ means direction perpendicular to the incident beam [15]. Thus, the maximum detectable poloidal wave-number varies with the radial positon of measuring chord. It decreases from the center to edge. It should be note that all discharges presented here are Ohmically heated hydrogen plasmas by gas-puffing fueling. And the minor radius $a$ is set at $0.255m$.

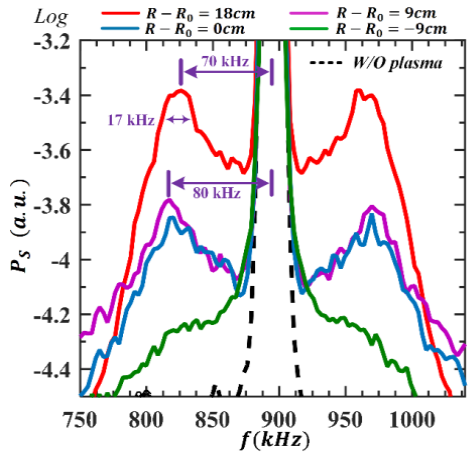

Figure 1. FCS spectra in the J-TEXT ohmic discharge ($I_P = 180kA, B_t = 2.0T, \bar{n}_{e0} = 3 \times 10^{19} m^{-3}$). The QC density fluctuations are seen on chords of $R - R_0 = 18cm$ and $R - R_0 = 0cm$, but not seen on $R - R_0 = -9cm$.

The typical density fluctuations spectra measured by FCS is shown in figure 1, which are from J-TEXT Ohmic discharge with parameters as: $I_P = 180kA, B_t = 2.0T, \bar{n}_{e0} = 3 \times 10^{19} m^{-3}$. The center frequency ($885 - 900 kHz$) with large amplitude is the intermediate frequency (IF) signal for measuring Faraday rotation angle. The two parts beside IF alike wings are the collective scattering signals, which are contributed by density fluctuations. There is no scattering signal before discharge, as the dashed line shows. The frequency difference between the IF ($f_0$) and scattering signal ($f_s$) is just the frequency of electron density fluctuation. The scattering spectrum should be symmetric relative to $f_0$, because the two probe beams are collinear combined. As Fig. 1 shows, the FCS spectra on the $R - R_0 = 18cm$ and $R - R_0 = 0cm$ chords display two peaks at frequencies of $|f_s - f_0| \cong 70\ \&\ 80 kHz$ respectively. And that frequency peaks have broad-band ($\Delta f$) about $17kHz$. The characteristic frequency of $\Delta f/f \cong 0.25$ indicates that the density fluctuations have QC features. Additionally, the mid-frequencies of that QC modes have a decreasing tendency with plasma minor radius. That is consistent with the reflectometer observations [14]. Furthermore, the QC mode is absent on the $R - R_0 = -9cm$ chord while it is noticeable at $R - R_0 = 9cm$. It shows the clear LFS/HFS asymmetry of the QC modes, which is the same to the QC mode observations on TEXTOR [12] and T-10 [11]. In other words, the QC modes are ballooned in the LFS. Actually, for the discharge in Fig. 1, the QC modes can be observed on all the measuring chords from $R - R_0 = 18cm$ to $R - R_0 = -6cm$. As mentioned above, the wave number of density fluctuations measured by FCS is limited as $k_\perp < 1.5 cm^{-1}$. Supposing that the electron temperature ($T_e$) varies from $800eV$ to $200eV$ while radial position increase from $r = 0cm$ to $r = 18cm$, then the normalized wave-number ($\rho_s k_\theta$) for the QC mode is estimated as $\rho_s k_\theta < 0.3$ in central and $\rho_s k_\theta < 0.1$ at edge ($r = 18cm$).

According to the collective scattering principle, the heterodyne detection using a twin-frequency source (one acts as local beam) is available to measure the propagation direction of density fluctuation wave at laboratory frame [18]. Of course, it demands the detector receiving the scattering wave from a particular direction. Thus, the frequency shift direction relative to incident beam is corresponding to the propagation direction. For normal far-forward scattering, it is almost impossible to discriminate the propagation direction of density fluctuations, because the scattering waves with positive and negative frequency shift are symmetric relative to the collection optical path of detector. But if the probe beam deviates from the optic axis of detector collection path, it will be available to identify the propagation direction. For FCS based on POLARIS, benefits from probe beam refracting by plasma, that deviation exists naturally. So if the refraction is enough significant, we could discriminate the propagation direction of that QC mode, by analyzing the heterodyne signal between local and scattering beams.

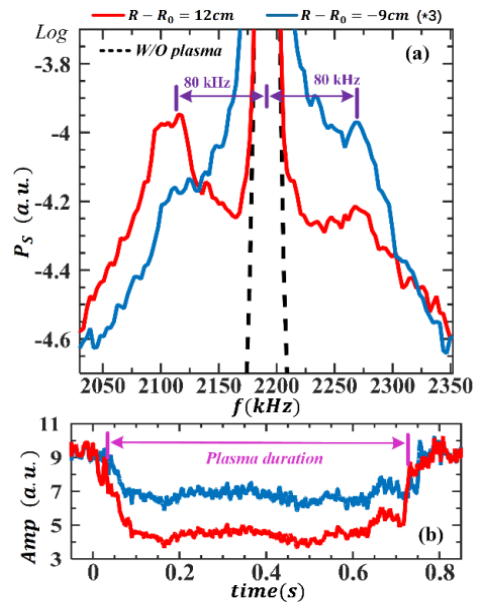

Figure 2. (a) Heterodyne FCS spectra at $R - R_0 = 12cm$ and $R - R_0 = -9cm$ for J-TEXT Ohmic discharge ($I_P = 180kA, B_t = 2.0T, \bar{n}_{e0} = 2 \times 10^{19} m^{-3}$). And the spectrum at $R - R_0 = -9cm$ is multiplied by three. (b) Time traces of the amplitudes of IF in top panel.

The heterodyne FCS spectra mixed by local and scattering beams are plotted in figure 2. The IF ($2175 - 2210 kHz$) is set for measuring electron density. As the

Fig. 2 (b) shows, the IF amplitude decreases 25% and 50% at $R - R_0 = 12cm$ and $R - R_0 = -9cm$ respectively, which results from the refraction of probe beams. As predicted above, the FCS spectra show obvious asymmetry relative to IF (Fig. 2 (a)). In this discharge, the frequency of local beam is set to be larger than probe beam, and the toroidal field is anticlockwise from the top view. Therefore, for channel at LFS ($R - R_0 = 12cm$), the negative frequency shift means the density fluctuations propagating in the ion diamagnetic direction at Lab. frame. And that is opposite for channel at HFS ($R - R_0 = -9cm$). In figure 2, both FCS spectra at LFS and HFS indicate that the QC mode propagates in ion direction at Lab. frame. Calculating with $k_\theta < 1 cm^{-1}$ and $f = 70 kHz$, the phase velocity of the QC mode at Lab. Frame is given as $v_{QC} > 4.4 km/s$. Considering the plasma $E \times B$ equilibrium flow usually rotates in electron direction, the QC velocity is underestimated at plasma frame. Therefore, it can be affirmed that the QC mode propagates in ion direction at plasma frame. In other words, the QC mode is an ion mode. In addition, the FCS spectrum at $R - R_0 = -9cm$ in Fig. 2 is multiplied by three. It means that the density fluctuations at HFS is much smaller than that at LFS.

As mentioned above, expect for the LF QC mode (70~120kHz), experiments in T-10 tokamak found another HF QC modes (150~250kHz) [11]. Actually in J-TEXT tokamak, the FCS also measured another QC mode whose characteristic frequency is higher than that ion QC mode. The heterodyne FCS spectra contain two different QC modes are showed in figure 3. As same to the spectra in Fig. 2, the ion QC mode (~75kHz) is clearly observed on both chords of $R - R_0 = 12cm$ and $R - R_0 = -3cm$. But the different and important thing is that another QC mode with characteristic frequency near 170kHz simultaneously appears at the $R - R_0 = -3cm$ spectrum. And the frequency shift direction for the HF QC is opposite to the LF QC. Thus, it can be easily deduced that this HF QC mode propagates in electron direction in Lab. Frame. Additionally, the HF QC mode is distinct at $r = 3cm$ chord but almost disappears at $r = 6cm$ (not shown in Fig. 3). It is mostly because the wave-number of HF QC mode falls some value between the measuring limitation of $r = 3cm$ and $r = 6cm$. Then its poloidal wave-number is estimated in the range of $k_\theta \cong 1.4 - 1.5 cm^{-1}$. Supposing $k_\theta = 1.45 cm^{-1}$ and $f = 170 kHz$, it gives the HF QC mode phase velocity at Lab. frame $v_{HFQC} \cong 7.3 km/s$ in electron diamagnetic direction. The plasma equilibrium flow poloidal velocity is estimated as $1 - 2 km/s$, inferred from carbon ion poloidal velocity measured by high-resolution spectrometer system [19]. Thus, the HF QC mode is inferred as electron mode.

In conclusion, the FCS measurement on J-TEXT has observed two different QC density waves, and they propagate in ion and electron direction respectively. The ion and electron QC modes have characteristic frequency of 50~100kHz and 150~200kHz respectively. Also, the typical wave-number for ion mode is limited by $\rho_s k_\theta < 0.3$ at central and $\rho_s k_\theta < 0.1$ for edge region, while that for the electron mode is estimated as $0.15 < \rho_s k_\theta < 0.3$. For Ohmic L-mode tokamak plasmas, the most unstable micro-instabilities with long wave-length ($\rho_s k_\theta < 1$) are predicted to be ITG and TEM modes. Also, the normalized wave-number for the two QC modes is closed to the theoretical predictions [20]. Take into account the propagation direction, it is reasonable to conclude that the ion QC mode is ITG instability and the electron QC mode is TEM instability. In addition, we should note that the TEM mode is difficult to measure, because its wave-number is close to the limitation of the FCS measurement on J-TEXT. Thus, here mainly studies the ITG mode in the following.

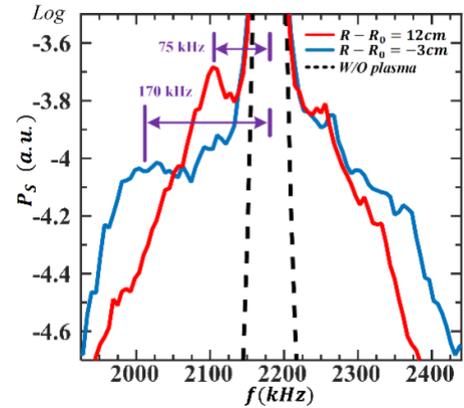

Figure 3. Heterodyne FCS spectra at $r = 12cm$ and $r = -3cm$ for J-TEXT Ohmic discharge. Both the QC ion mode and another QC electron mode are observed.

As suggested that the stability of ITG mode is strongly related to plasma density and confinement saturation in tokamaks, we have studied the behaviors of ITG mode during density ramp-up. For the discharge using continual gas-puffing to raise density in one shot, the FCS spectra ($r = 12cm$) evolution with central averaged electron density ($\bar{n}_{e0}$) is plotted in Fig. 4. The discharge parameters are $I_P = 180 kA, B_t = 2.0T$, and the density climbs from $1.5 \times 10^{19} m^{-3}$ to $4.5 \times 10^{19} m^{-3}$. It is need to note that the FCS spectra in Fig. 4 is normalized by density ($P_{FCS}/\bar{n}_{e0}$). According to the behaviors of ITG modes, this discharge can be divided into three density regimes. In the low density (LD) range ($\bar{n}_{e0} < 2 \times 10^{19} m^{-3}$), the ITG mode is too weak to be observed, as the spectrum for $\bar{n}_{e0} = 1.8$ shows. In the medium density (MD) range ($\bar{n}_{e0} = 2 - 3.5 \times 10^{19} m^{-3}$), the ITG mode is noticeable and enhances slowly with density increase. Meanwhile its characteristic frequency almost keeps constant. In the high density (HD) range ($\bar{n}_{e0} > 3.5 \times 10^{19} m^{-3}$), the amplitude of ITG mode increases substantially with density, and its characteristic frequency decreases simultaneously. As Fig. 4 shows, during the HD regime, the normalized fluctuation power ($P_s/\bar{n}_{e0}$) for ITG mode has trebled, and its central frequency decreases from $80 kHz$ to $40 kHz$. In conclusion, the ITG mode behaviors have two bifurcation points. The first point is the appearance of ITG mode, and the corresponding critical density is $\sim 2 \times 10^{19} m^{-3}$. The second point is the abrupt amplitude increase and frequency decrease for ITG mode, and the corresponding critical density is $\sim 3.5 \times 10^{19} m^{-3}$. It is believed that the SOC regime is related to the bifurcation behavior of ITG mode. According to the empirical scaling by *Shimomura* [21], the critical density for SOC is predicted as $n_e^c = I_p \mu_0 \sqrt{A_i}/(2\sqrt{2}\pi a^2) \cong 3.9 \times 10^{19} m^{-3}$. It is much close to the second bifurcation point.

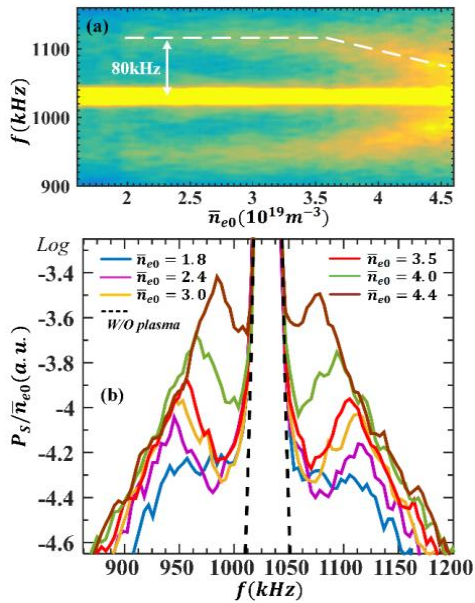

Figure 4. Evolution of normalized FCS spectra ($R - R_0 = 12 cm$) with the central averaged electron density ($\bar{n}_{e0}$) for a density ramp-up discharge. The discharge parameters are $I_P = 180 kA, B_t = 2.0 T$.

Furthermore, we have studied the ion temperature profile and the ITG critical parameters $\eta_i = L_n/L_{Ti}$ for slab branch and $R_0/L_{Ti}$ for toroidal branch. The ion temperaure profile at edge ($r > 0.6a$) is given by high-resolution spectrometer system, and plasma density profile is obtained by POLARIS [22]. As Fig. 5(c) shows, both $\eta_i$ and $R_0/L_{Ti}$ (at $r = 0.7a \sim 18 cm$) increase with density climbing. That is why ITG mode enhances with density. The correlation between the ITG mode amplitude and $\eta_i$ has been affirmed by experiments in CLM [23]. Also, the critical value of $\eta_i$ for ITG mode occurrence (where $\bar{n}_{e0} \approx 2 \times 10^{19} m^{-3}$) is about 1.5. It is consistent with the theoretical prediction. In addition, the ion temperature (Fig. 5(a)) at core region ($0.6a < r < 0.8a$) shows abrupt declining after the density exceeds $3.5 \times 10^{19} m^{-3}$. Meanwhile, the density profile peaking factor ($n_{e0}/\bar{n}_{eo}$) and ion energy in the region ($r > 0.6a$) both reaches a maxima, as Fig. 5(b) indicates. That implicates the enhancement of ion energy losses and global confinement degradation. And the critical density $\bar{n}_{e0} \cong 3.5 \times 10^{19} m^{-3}$ is highly consistent with the threshold of the abrupt enhencement for ITG mode.

In summary, the ITG and TEM instabilities with quasi-coherent spectra have been affirmed experimentally for the first time, by the far-forward collective scattering measurement in J-TEXT tokamak. The ITG mode shows a bifurcation behavior after plasma density exceeds a critical value, where it enhances substantially. At the same point, the ion energy loss increase and global confinement degradation are also observed. It gives the direct experimental evidence for the ion thermal transport driven by ITG mode.

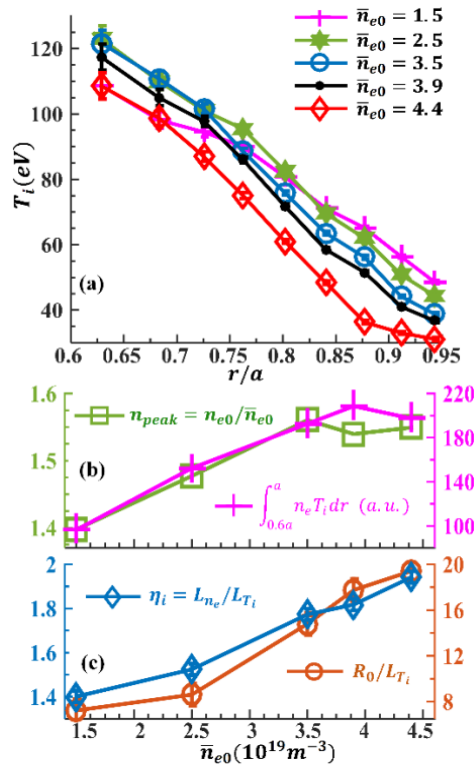

Figure 5. Dependences of ion temperature profile (a), density profile peaking factor and ion energy in edge region (b), $\eta_i = L_n/L_T$ and $R_0/L_T$ at $r/a = 0.7$ (c) on the central line-averaged electron density.

**Acknowledgement**: This work was supported by the National Natural Science Foundation of China under Grant Nos. 0204131240 and 11575067.